\documentclass[prb,a4paper,showpacs,superscriptaddress,twocolumn]{revtex4}
\usepackage{amssymb}
\usepackage{amsmath}
\usepackage{amsfonts}
\usepackage{graphicx}
\usepackage{bm}
\usepackage{color}
\usepackage{float}
\usepackage{verbatim}
\usepackage{asymptote}
\usepackage{epstopdf}

\definecolor{DarkBlue}{rgb}{0,0,0.6}

\def\ve{\varepsilon}

\def\vf{\varphi}
\def\de{\partial}

\def\mT{\mathcal{T}}
\def\mR{\mathcal{R}}

\def\psiu{\psi_\uparrow}
\def\psid{\psi_\downarrow}

\def\br{\mathbf{r}}

\def\bp{\mathbf{p}}

\def\fv{v_\textrm{F}}

 1

\begin{document}

\title{Floquet spectrum and driven conductance in Dirac materials: \\ Effects of Landau--Zener--St\"uckelberg--Majorana interferometry}

\author{Ya.~I.~Rodionov} \affiliation{Institute for Theoretical
and Applied Electrodynamics, Russian Academy of Sciences,
Izhorskaya str. 13, Moscow, 125412 Russia}
\affiliation{National University of Science and Technology MISIS, Moscow, 119049 Russia}

\author{K.~I.~Kugel} \affiliation{Institute for Theoretical and
Applied Electrodynamics, Russian Academy of Sciences, Izhorskaya
str. 13, Moscow, 125412 Russia}
\affiliation{National Research University Higher School of Economics, Moscow, 109028 Russia}

\author{Franco~Nori} \affiliation{Center for Emergent Matter
Science, RIKEN, Wako-shi, Saitama, 351-0198, Japan} \affiliation{Department
of Physics, University of Michigan, Ann Arbor, MI 48109-1040, USA}

\begin{abstract}
Using the Landau--Zener--St\"uckelberg--Majorana-type (LZSM) semiclassical approach, we study both graphene and a thin film of a Weyl semimetal subjected to a strong AC electromagnetic field. The spectrum of quasi energies in the Weyl semimetal turns out to be similar to that of a graphene sheet.
Earlier it has been predicted qualitatively that the transport properties of strongly-irradiated graphene oscillate as a function of the radiation intensity [S.V. Syzranov \textit{et al.}, Phys. Rev. B {\bf 88}, 241112 (2013)].
Here we obtain rigorous quantitative results for a driven linear conductance of  graphene and a thin film of a Weyl semimetal.
The exact quantitative structure of oscillations exhibits two contributions.
The first one is a manifestation of the Ramsauer--Townsend effect, while the second contribution is a consequence of the LZSM interference defining the spectrum of quasienergies.
\end{abstract}


\date{\today}

\maketitle

\section{Introduction}
Graphene nanoribbons, superlattices and other mesoscopic graphene-based structures attract considerable current
interest.~\cite{RozhkovNoriPhRep2011,Kusmartsev2013,RozhkovPhPep2016} Size effects in such systems allow for the fine tuning of their electronic spectra and, as a result, manipulating their transport and optical characteristics. Periodic superstructures are of special importance since the periodicity gives rise to additional features in the electronic band structure, such as opening band gaps and forming new Dirac points.~\cite{SyzranovPRB2008,FertigPRL2011,SRKN} Unfortunately, it is not an easy task to create graphene superlattices and their tunability is rather limited. However, it is well known that in quantum mechanics, there exists a profound similarity between the effects of spatial and temporal periodicity. Indeed, an analog of the Bloch theorem (namely, the Floquet theorem) also works for systems in time-periodic fields. Namely, the particle energy should become a quasienergy $\ve$ bounded within its Floquet zone $-\hbar\omega_0/2 < \ve < \hbar\omega_0/2$, where $\hbar$ is the Planck's constant and $\omega_0$ is the characteristic frequency of the uniform field. The quasienergy spectrum can exhibit minigaps dependent on the amplitude of the field.~\cite{Silveri_PRB2013} The  concept of quasienergy was first introduced in atomic physics in the seminal papers by Zel'dovich~\cite{ZeldovichJETP1967} and Ritus~\cite{RitusJETP1967} and was widely used in different fields of physics, especially at the nanoscale (see the review article in Ref.~\onlinecite{KohlerPhRep2005} and references therein).

Recently, systems with a Dirac Hamiltonian driven by a periodic external electromagnetic field are attracting considerable interest.~\cite{GanichevPhRep2014,SavelAlexPRB2011,PerezPiskPRB2014,
PerezPiskPRA2015,kibis1,FertigPRL2016} For example, the quasienergy concept has been implemented for graphene interacting with the electromagnetic field.~\cite{FertigPRL2011,OkaAokiPRB2009,ZhouWu2011,BuslPRB2012}
For the case of graphene, a profound analogy between the spatial and temporal modulation (special-temporal duality) is discussed in detail in Ref.~\onlinecite{SavelAlexPRB2011}.
However, most of these studies have dealt either with  the perturbative response of graphene in  weak electromagnetic time-dependent fields or resorted to numerical analysis.

Meanwhile, the effects related to the minigaps in the quasienergy spectrum should become even more pronounced when increasing the field amplitude. Indeed, as was proved in Ref.~\onlinecite{SyzranovPRB2008}, the actual energy of carriers exhibits gaps proportional to the perturbatively small
($eE_0\fv\ll\hbar\omega_0^2$) amplitude of a periodic field. It is therefore important to study analytically the spectrum in the opposite limit of strong fields. This limit corresponds to a semiclassical description in the time domain. Being described by a two-component wave function, graphene is very akin to a two-level system. In fact, it is a good realization of a Landau--Zener interferometer~\cite{FeigelmanEJPB2003,MakhlinJLTP2007,
Shevchenko_PRB2012, Satanin_PRB2012,Sarma_PRL2013,Satanin_PRB2014, Fillion_arxiv2016, Fillion_arxiv2016a} with the range of applicability growing with the field amplitude (see also a detailed review article Ref.~\onlinecite{NoriPhRep2010} and references therein). It is relevant to mention the four seminal papers on this subject, namely those of Landau~\cite{Landau1932}, Zener~\cite{Zener1932}, St\"uckelberg~\cite{Stuckelberg1932}, and Majorana~\cite{Majorana1932}; so, hereafter, we use the terms Landau--Zener--St\"uckelberg--Majorana (LZSM) transitions or interferometry.  Here, we are dealing with the interference of the wave functions corresponding to multiple transitions between the electron states.

In the present paper, we focus on the specific features of the conductance of a Dirac material driven by the incident electromagnetic wave.
Earlier it was predicted that the conductance of a graphene p-n junction of  strongly-irradiated graphene oscillates as a function of the radiation intensity.~\cite{SRKN} However, the technique used in Ref.~\onlinecite{SRKN} allows making just a qualitative prediction of the oscillations amplitude.
 Here we obtain a closed analytical expression for the linear driven conductance. We show that the result is also applicable to another Dirac material, namely, to a thin film of a Weyl semimetal, which is a three-dimensional analog of graphene.~\cite{Vafek_AnnRevCondMat2014,Turner_arxiv2013} We demonstrate that the aforementioned features are also reproduced by Weyl semimetals.

\begin{figure}[t] \centering
\includegraphics[width=0.45\textwidth]{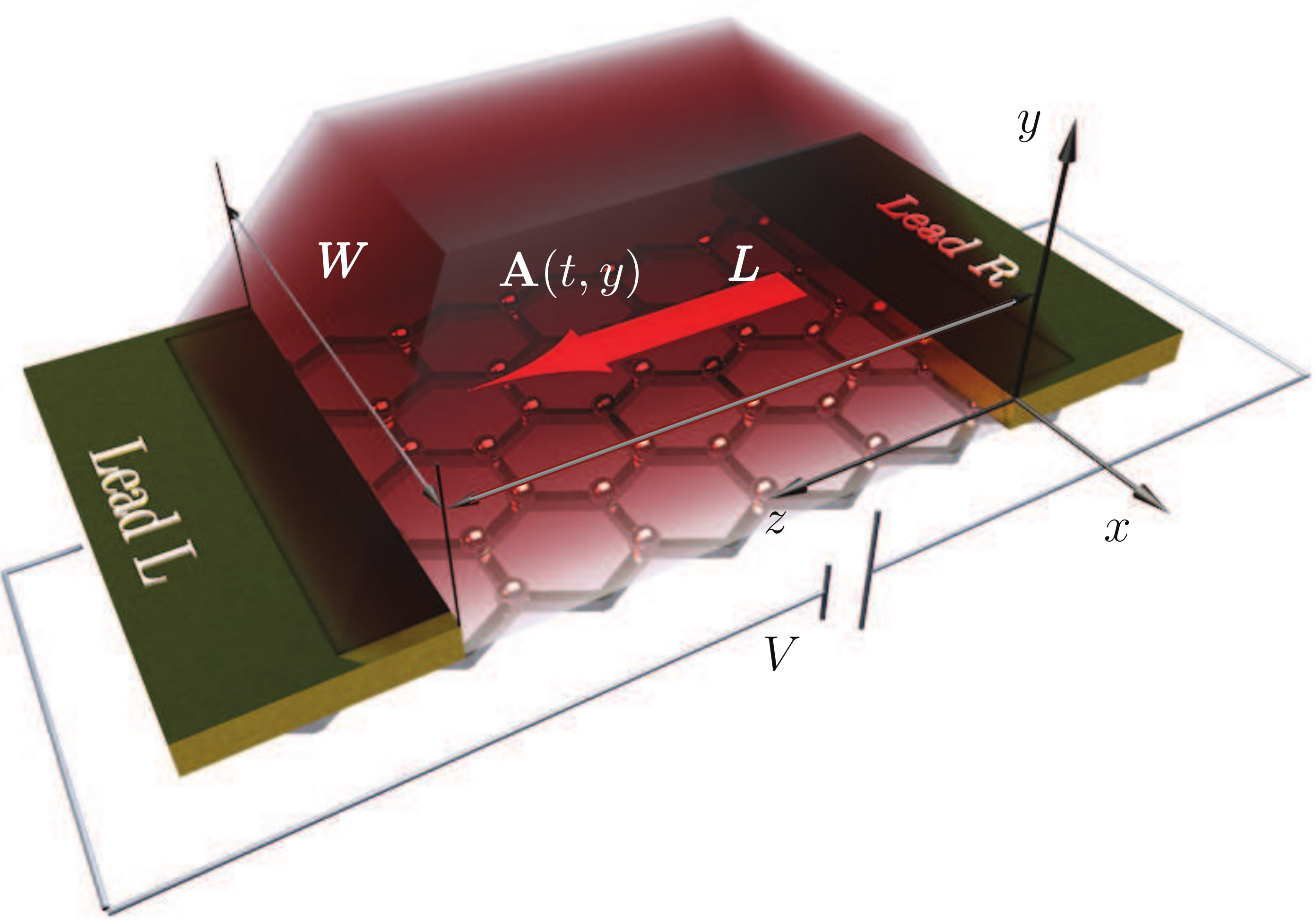}
\caption{(Color online)
     \label{strip2}
Schematics of a set-up for graphene or a thin film of the Weyl semimetal. A normal-incident linearly-polarized electromagnetic wave irradiates the graphene strip. The red arrow shows the direction of the polarization of the vector potential $A(t,y)$. The illuminated rectangular area of the sheet is shaped by two metallic leads (electron reservoirs). The thickness of the leads is assumed to be much larger than the skin-effect depth. The skin effect protects the reservoir electrons from the effect of irradiation.  The linear response of the system is measured by applying an infinitesimal bias voltage $V$.
          }
\end{figure}

We consider a  graphene sheet or a thin layer of a Weyl semimetal subjected to a strong normally-incident linearly-polarized AC electromagnetic field (see Fig.~\ref{strip2}). As mentioned above, the wave function of a periodically-driven system satisfies the Floquet theorem
\begin{gather}
  \label{fl1}
   \psi_\alpha(t+T)=\exp{(-i\ve_\alpha T)}\psi_\alpha(t),
\end{gather}
where the subscript $\alpha$ enumerates the states, $T=2\pi/\omega_0$ is the period of the driving field, and $\ve_\alpha$ is a so-called quasienergy. The quasienergy plays a role similar to the crystal momentum in a spatially-periodic system. The state described by the wave function in ~\eqref{fl1} is referred to as a Floquet state.

A problem is that the quasienergy does not correspond to any stationary state (in contrast to the crystal momentum). A Floquet state~\eqref{fl1} with quasienergy $\ve_\alpha$ is, in fact, a linear combination of all possible  modes with energies $\ve_\alpha+n\omega_0$, where $n$ is an integer. To relate it to some quasiparticle state, one has to think of a lifetime of such a state and its stationary distribution function, if there is any.

It was demonstrated earlier, e.g. in Ref.~\onlinecite{SRKN} (for the case of a strong driving field) and later in Ref.~\onlinecite{kibis1}, that the quasienergy spectrum $\ve(\bp)$ becomes highly anisotropic and forms a set of additional Dirac points in momentum space. Even more, due to the interference of two successive LZSM transitions, one of the Fermi velocities acquires a non-trivial oscillating dependence on the driving field intensity. Reference~\onlinecite{kibis1} goes as far as to argue that one can compute the DC conductivity using the quasienergy spectrum as an effective spectrum of charge carriers, hence, employing Fermi-liquid type expressions with stationary Fermi distributions. This prediction is indeed a very tempting one to make. However, such treatment overlooks the evident non-equilibrium dynamics of the quasienergy states which is essential for the accurate description of the transport phenomena. It also takes into account the contribution from just one Dirac point (the addition of other Dirac points lead to a divergent answer).

Contemplating such state of affairs, we have to ask an inevitable question: is it possible to construct a quantum mechanical observable, which is directly related to the quasienergy? The answer we give in our paper is positive. The observable in question is the so-called driven conductance. It can be realized in the following geometry (see Fig.~\ref{strip2})
The graphene stripe (with length $L$ and width $W$) is irradiated by a linearly-polarized electromagnetic wave at normal incidence. The corresponding vector potential in the plane of the graphene sheet is
\begin{gather}
   \label{vector}
 \mathbf{A}(t)=(0,0,(cU_0/e)\sin\omega_0 t),
\end{gather}
where the electric field amplitude $E_0=U_0\omega_0/e$, and $e$ is the electron charge.

In this paper, we limit ourselves to the study of the ballistic regime for the electron transport in the irradiated Dirac material. In such limit, the effects related to the Floquet spectrum are the most clearly pronounced. The left and right edges of a ballistic graphene sheet are connected to equilibrium electron reservoirs, which screen the corresponding parts of the sheet from the incident electromagnetic field due to the skin effect. An infinitesimal bias voltage $V$ is created by the chemical potential difference in the left and right reservoirs, $\mu$ and $\mu+V$, respectively. The linear response of the system is characterized by the electric current averaged over the period $T$ of the electromagnetic oscillations.
The driven conductance is defined as
\begin{gather}
   G=\frac{\bar{I}}{V}=\frac{\frac{1}{T}\int_0^T I(t)dt}{V}.
\end{gather}
In ballistic graphene,  the thermalization of electrons takes place only in the electron reservoirs. Therefore, the electrons inside the sheet are highly overheated. We compute the conductance in the limit
\begin{gather}
\label{cond1}
\kappa=\frac{\hbar\omega_0}{\fv U_0}\ll1,\\
\label{cond2}
\beta=\frac{\mu}{\hbar\omega_0}\ll\kappa,
\end{gather}
where $\fv$ is  the Fermi velocity.
Condition~\eqref{cond1} corresponds to a strong resonant interaction between the charge carriers and the applied electromagnetic field; while condition~\eqref{cond2} allows for an analytical solution of the conductance problem. The  optimum frequency lies in the THz range (see Section IV for details). One of the results of Ref.~\onlinecite{SRKN} is that, in the limit~\eqref{cond1}, the quasienergy spectrum can be computed analytically in the vicinity of each Dirac point. The spectrum near the Dirac points lying on the $p_z$ axis is of particular significance for future analysis
\begin{gather}
  \label{quasi-en0}
   \begin{split}
   \ve_{p,n}&=\pm \sqrt{v_{z{\rm F}}^2\delta p_z^2+v_{x{\rm F}}^2(U_0)p_x^2},\\
   v_{z{\rm F}}&=\fv,\\
   v_{x{\rm F}}(U_0)&=\fv\sqrt{\frac{\hbar\omega_0}{\pi \fv U_0}}\,\left|\sin\left(\frac{2U_0\fv\lambda}{\hbar\omega_0}{+}
   \frac{\pi}{4}\right)\right|,\\
   \delta p_z&=p_z-\frac{\hbar\omega_0 n}{\fv},
  \end{split}
\end{gather}
where
\begin{gather}
\lambda=\sqrt{1-\frac{p_z^2}{U_0^2}}+\frac{p_z}{U_0}
\arcsin\frac{p_z}{U_0}.
\end{gather}
Here $n$ is an integer numbering a Dirac point. The quasienergy spectra of graphene and a thin film of Weyl semimetal turn out to be identical. They are related to each other by the change of variables $p_x\rightarrow p_\bot=\sqrt{p_x^2+p_y^2}$.

The results of this paper can be summarized as follows. The conductance of graphene and a thin film of Weyl semimetal is obtained analytically in the limit~\eqref{cond1}-\eqref{cond2}. In the simplest case, $L\ll \fv\hbar/\mu$, it is represented by a sum of three contributions of different nature
\begin{gather}
\label{result}
 \begin{split}
   G(\omega_0,U_0)&=G_{\rm I}+G_{\rm R}(\omega_0,U_0)+G_{\rm F}(\omega_0,U_0),\\
   G_{\rm I}&=C_1\mu^{2} ,\\
   G_{\rm R}(\omega_0,U_0)&=C_2\mu^{2} J_0^2\left(\frac{2v_{\rm F}U_0}{\hbar\omega_0}\sin\frac{\omega_0 L}{2v_{\rm F}}\right)(2\cos\frac{2\mu L}{\hbar v_{\rm F}}+1),\\
   G_{\rm F}(\omega_0,U_0)&=C_3\mu^{2}\frac{v^2_{x\rm F}}{\fv^2}\sin^2\frac{\mu L}{\hbar v_{\rm F}},
    \end{split}
\end{gather}
where $J_0(x)$ is the Bessel function of zeroth order and $C_{1,2,3}$ are numerical constants computed below. Here, $G_{\rm I}$ is the non-oscillatory part of the conductance and is of no interest to us. The second term of $G$, $G_{\rm R}$, reveals oscillations of the conductance as a function of the driving field amplitude $U_0$ (a prefactor of the sine in the argument) \textit{as well as} the driving-field frequency $\omega_0$ (in the sine argument). This term can be considered as a manifestation of the Ramsauer--Townsend effect.~\cite{Griffits}
%~\cite{Bohm_book}
Indeed, we will see that this term stems from the quantum interference between the incident and reflected components of the quasiparticle wave functions.

The third term $G_{\rm F}$ is associated with the LZSM physics and Floquet excitations. It is proportional to the field-dependent velocity of a Floquet excitation $v_{x\rm F}$. Experimentally, one can  separate the most interesting dependence of the Floquet excitation velocity on the external field amplitude by measuring the conductance at the specific frequency
\begin{gather}
 \label{omega}
\omega_k=\frac{2v_{\rm F}}{L}\arcsin\frac{\hbar\omega_k x_k}{2v_{\rm F}U_0},
\end{gather}
where $x_k$ is the $k$th zero of the Bessel function $J_0$. At these frequencies, the contribution of the second term in~\eqref{result} is excluded and the conductance is simply $G_{\rm I}+G_{\rm F}$.
We have
\begin{gather}
   v^2_{x\rm F}(U_0)\sim G(\omega_k,U_0)-G\left(\omega_k,U_n\right),
\end{gather}
where $U_n=\hbar\omega_0\pi(n-1/4)/2v_{\rm F}$.

The paper is organized as follows. Section~\ref{floq-spec} presents the general formalism and spectrum of 2D and 3D Dirac materials subject to an incident electromagnetic wave. In Section~\ref{conduct}, we calculate the driven conductance in the 2D and 3D cases. Concluding remarks are given in Section IV. Some technical issues are discussed in the Appendix.

\section{Floquet spectrum\label{floq-spec}}

\subsection{Hamiltonian}

Throughout this paper, we use the atomic units $\hbar=\fv=e=1$. In some key places, the usual units are restored. The Hamiltonian describing quasiparticles in the vicinity of Dirac point has the form
\begin{gather}
 H=\bm{\sigma}\cdot\bp,\ \
\end{gather}
where $\bm{\sigma}$ is the triad of Pauli matrices in the space of the two sublattices in graphene.

For graphene, we use the $xz$ basis, which is rotated with respect to the standard one (see below). In the 3D case, the thickness of a sample in the $y$ direction is assumed to be much smaller than the skin depth of the material, $D\ll c\sqrt{\rho\ve_0/(\omega_0\mu_r)}$, where $\rho$ is the electrical resistivity of the  semimetal, $\mu_r$ is its relative permeability, and $\ve_0$ is the permittivity of vacuum. The typical Weyl semimetal Cd$_2$As$_3$ has resistivity $\rho\gtrsim 10^5$ n$\rm\Omega\cdot cm$ at temperatures $T\gtrsim 200$ K (see Ref.~\onlinecite{Liang2015}) and the corresponding skin depth is $\gtrsim 0.1\, \mu\rm m$ at $\mu_r\sim1$, and $\omega_0\sim 1$ THz. The polarization of an applied field is linear (in $z$ direction) and given by vector potential~\eqref{vector}. The Hamiltonian within the irradiated region is
\begin{gather}
  \label{ham1}
  \begin{split}
   H(t)=\sum\limits_{i}\sigma_ip_i-U(t)\sigma_z\, ,
  \end{split}
\end{gather}
where $U(t)=U_0\sin\omega_0 t$. In 3D, the small thickness in the $y$ direction allows neglecting the $y$ dependence of the vector potential.

This means that the Hamiltonian commutes with the full vector of the momentum operator $\bp$: $[H,\bp]=0$ both in 2D and 3D. Hence, the momentum is conserved and we turn to the basis
\begin{gather}
  \Psi(t)=\psi_\bp(t)\exp(i\bp\br),
\end{gather}
in which the Hamiltonian takes the form
\begin{gather}
  \label{ham2}
  \begin{split}
  H_{2D}&=\begin{pmatrix}
        p_z-U(t)\ &\ p_x\\
        p_x\ &\ -(p_z-U(t))
   \end{pmatrix},\\
   H_{3D}&=\begin{pmatrix}
        p_z-U(t)\ &\ p_\bot e^{-i\chi}\\
        p_\bot e^{i\chi}\ &\ -(p_z-U(t))
   \end{pmatrix},\ \
  \end{split}
\end{gather}
where $\chi$ is the azimuthal angle: $p_x=p_\bot\cos\chi,\ p_y=p_\bot\sin\chi$. Similarly to the static case, the 2D and 3D Hamiltonians are related via the unitary transformation $U(\chi)=\cos\frac{\chi}{2}-i\sigma_z\sin\frac{\chi}{2}$.
Hence, they share the eigenfunctions
\begin{gather}
  \begin{split}
    H_{2D}=U^{-1}(\chi)H_{3D}U(\chi),\ \
   \psi_{3D}(t)=U(\chi)\psi_{2D}(t),
   \end{split}
\end{gather}
as well as eigenvalues. The latter are the quasienergies. Therefore, even in the presence of a linearly-polarized external field, the thin 3D and 2D Dirac materials retain their unitary equivalence. All the formulae derived for a Weyl semimetal are valid for graphene up to a geometrical prefactor. The spectrum of the quasienergies was derived in Ref.~\onlinecite{SRKN}. Since we are going to use the details of the spectrum in the rest of the paper, we briefly present the main steps. From now on, we consider the case of a Weyl semimetal. The graphene twin of each formula can be obtained by a simple replacement $\bp_\bot\rightarrow p_x$ and the change of the integration measure $d\chi \sin\vf d\vf\rightarrow d\vf$.

Due to condition~\eqref{cond1}, the evolution of the system is semiclassical and one immediately obtains the pair of wave functions (see Appendix for details)
\begin{gather}
  \label{gen-sol}
   \begin{split}
  \psi_\bp&=\chi_{\bp\pm}(t)\exp{\left(\pm i\int^t q \, dt\right)},\\
  \\
      \chi_{\bp\pm}&=\frac{1}{\sqrt{2}}\begin{pmatrix}
      \sqrt{\frac{q\pm(U-p_z)}{q}}\\
      \, \\
      \mp\sqrt{\frac{q\mp(U-p_z)}{q}}
      \end{pmatrix},
   \end{split}
\end{gather}
where
\begin{gather}
 \label{phase}
 q\equiv q(t)=\sqrt{[U(t)-p_z ]^2+p^2_\bot}.
\end{gather}
The square root defining semiclassical momentum~\eqref{phase} should be understood as an analytic function of time. Its regular branch is fixed by the condition $q>0$ when $U(t)>p_z$.  There, $q(t)$ has turning points at times satisfying the condition $[U(t)-p_z]^2+p_\bot^2=0$. These points give rise to the nonzero probability of a transition between wave function pair~\eqref{gen-sol}.

\begin{figure}[t]
  \includegraphics[width=85mm]{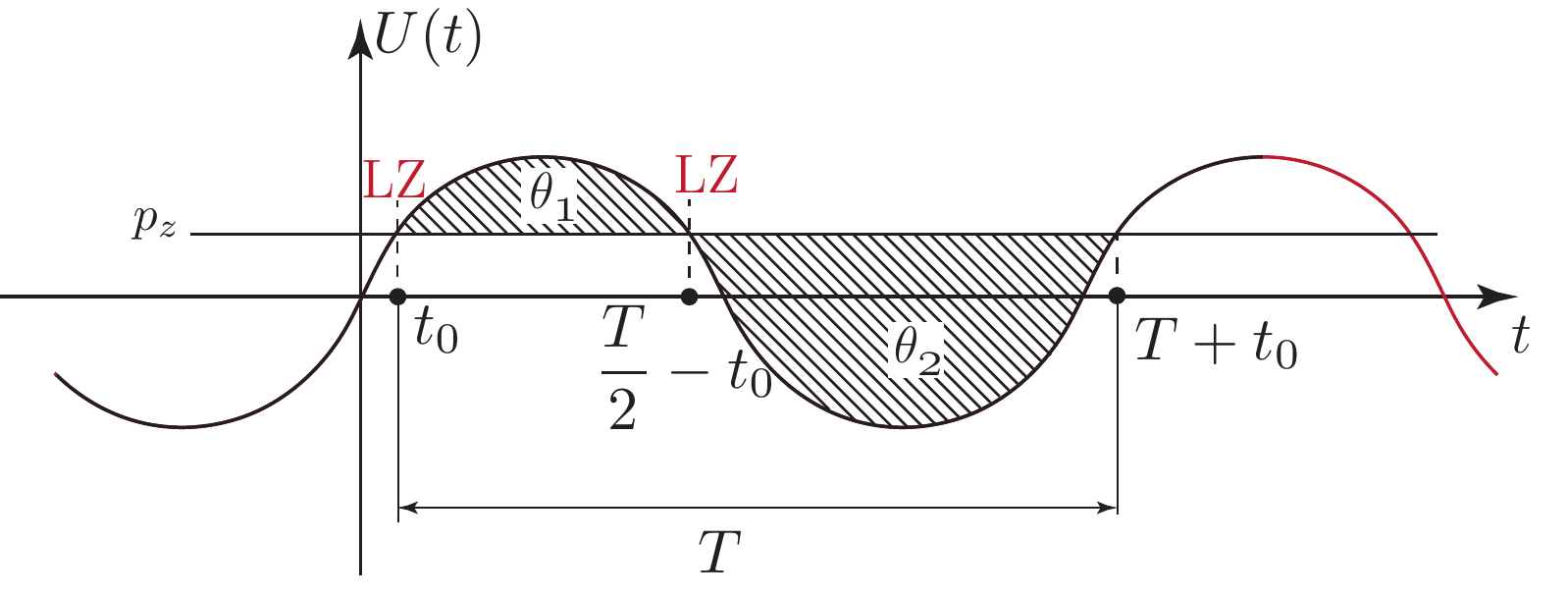}
  \caption{\label{graph1}
  (Color online) The interference of LZSM transitions in a Weyl semimetal, $t_0=\omega_0^{-1}\arcsin(p_z/U_0)$.
          }
\end{figure}

\begin{figure}[t]
\begin{center}
  \includegraphics[width=85mm]{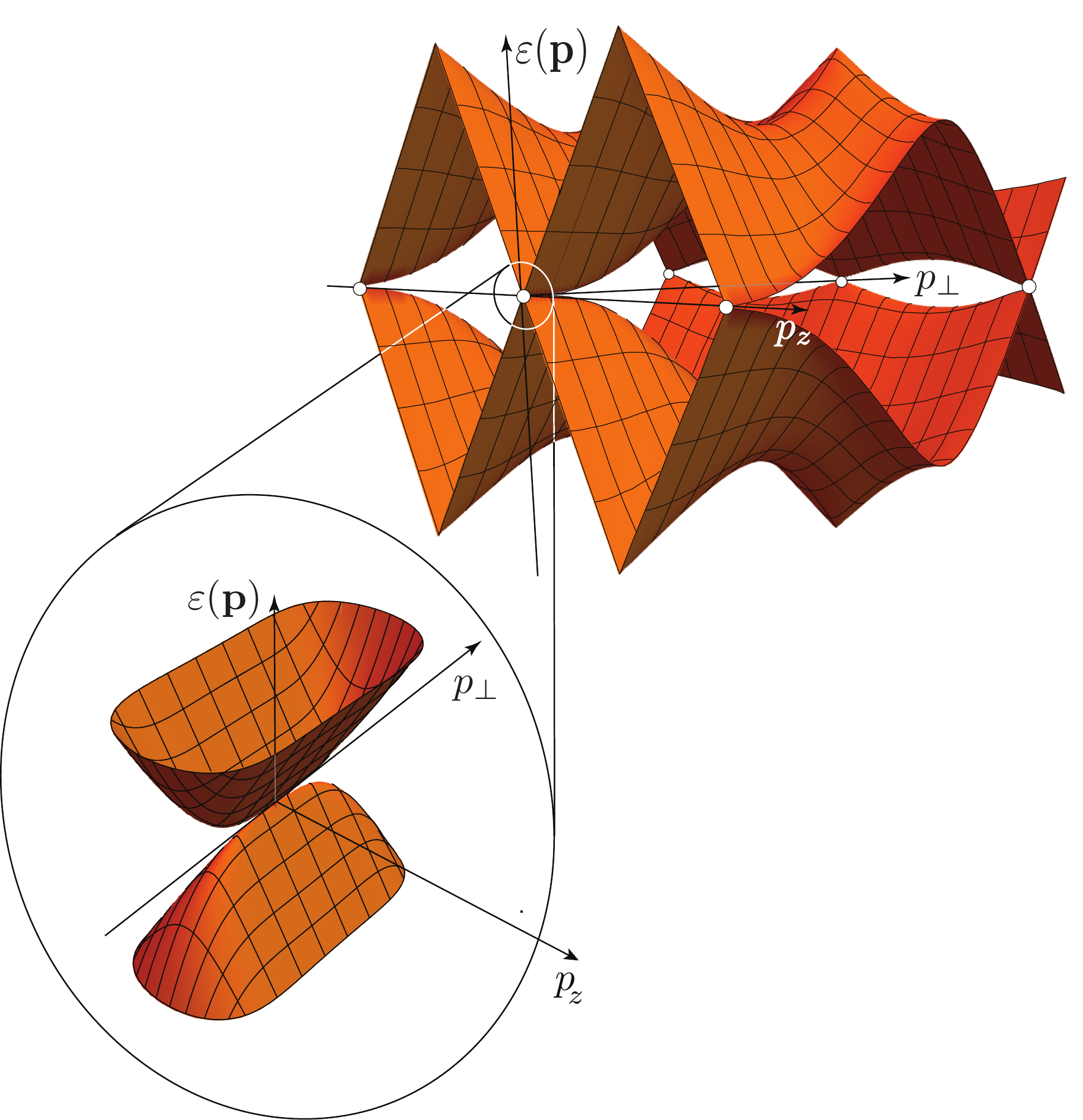}
  \end{center}
  \caption{\label{graphene-spectrum} (Color online) Resonant spectrum near the Dirac points at $U_0=15/8\omega_0$.
  { The asymmetric non-conical shape is clearly visible.}}
  \end{figure}

The case most relevant for our current task is that of a small transverse momentum
\begin{gather}
   \label{cond3a}
   p_\bot\ll U_0.
\end{gather}
It corresponds to the situation when transition point lies on the real axis: $U(t)-p_z=0$. In fact, this is precisely the case for a LZSM transition. The system undergoes two consecutive LZSM transitions (Fig.~\ref{graph1}) at times $t_0=\omega_0^{-1}\arcsin(p_z/U_0)$ and $T/2-t_0$. We note that condition~\eqref{cond3a} is not very restrictive, since the amplitude $U_0$ is still large.
One can also see that in limit~\eqref{cond3a}, the inequality $|U(t)-p_z|\gg p_\bot$ holds for all points, which are not too close to the time values corresponding to the LZSM transitions. This provides a significant simplification of eigenfunctions~\eqref{gen-sol}
\begin{gather}
   \chi_{\bp+}\approx\begin{pmatrix}
                        1\\
                        0
                     \end{pmatrix},\quad
    \chi_{\bp-}\approx\begin{pmatrix}
                        0\\
                        1
                     \end{pmatrix},\ \ p_\bot\ll U_0.
\end{gather}
Therefore, the evolution of the wave functions between the LZSM transitions can be represented by a simple diagonal operator
\begin{gather}
 \label{evol1}
   \mathcal{U}(t)=\begin{pmatrix}
               \exp{(i\int^tq\,dt)}\ &\ 0\\
                 0 \ &\  \exp{(-i\int^tq\,dt)}
           \end{pmatrix} .
\end{gather}
In this basis, the semiclassical scattering matrices associated with each transition have the standard form (see, e.g., Ref.~\onlinecite{lz-matrix})
\begin{gather}
 \label{lz-matrix}
  \begin{split}
   \mathcal{L}_{1,2}=\begin{pmatrix}
         \sqrt{P}\ &\ \sqrt{1-P}e^{\mp i\vf_s}\\
         \\
         -\sqrt{1-P}e^{\pm i\vf_s}\ &\  \sqrt{P}
       \end{pmatrix},
  \end{split}
\end{gather}
where the upper sign corresponds to the earlier transition, $P=\exp(-\pi\nu)$, $\vf_s=\pi/4-\nu/2\ln(\nu/2e)+\hbox{arg}\,\Gamma(i\nu/2)$, ($\nu=p_\bot^2/(U_0\omega_0)$ is the Stokes phase, and $\Gamma(x)$ is the Euler gamma function). Therefore, the evolution of a wave function during the period $T$ consists of two intervals: (1) $U(t)>p_z$  and (2) $U(t)<p_z$, divided by two LZSM transitions. The quasienergies are now defined by  relation~\eqref{fl1} as the eigenvalues of the evolution operator
\begin{gather}
  \label{evol-full}
  \mathcal{U}_T=\mathcal{U}\left(t_0+T,\frac{T}{2}-t_0\right)
  \mathcal{L}_2\mathcal{U}\left(\frac{T}{2}-t_0,t_0\right)\mathcal{L}_1 .
\end{gather}
Denoting the accumulated semiclassical phases entering~\eqref{evol1} as
\begin{gather}
  \label{phases2}
   \theta_{1}=\int_{t_0}^{\frac{T}{2}-t_0}|q|\,dt,\ \    \theta_{2}=-\int_{\frac{T}{2}-t_0}^{T+t_0}|q|\,dt,
\end{gather}
and taking the trace of the full evolution operator~\eqref{evol-full}, we arrive at the eigenvalue equation
\begin{gather}
  \label{eigen-val}
   \cos\ve_p T=P\cos(\theta_1+\theta_2)-(1-P)\cos(\theta_1-\theta_2-2\vf_s).
\end{gather}
The semiclassical phases $\theta_{1,2}$ were calculated in Ref.~\onlinecite{SRKN}
\begin{gather}
  \label{phases1}
  \begin{split}
   \theta_{1,2}&=\mp\frac{U_0}{\omega_0}\bigg[2\mp\frac{\pi p_z}{U_0}+\left(\frac{p_z}{U_0}\right)^2+\frac{p_\bot^2}{U_0^2}
   \ln\frac{4\sqrt{e}U_0}{|p_\bot|}\Bigg] .
  \end{split}
\end{gather}
With the help of~\eqref{phases1}, the solution of Eq.~\eqref{eigen-val} can be written explicitly in the most interesting case $\ve_\bp\ll\omega_0$. We expand all the values in~\eqref{eigen-val} up to second order in $p_{\bot},\ p_z$. The result is given by Eq.~\eqref{quasi-en0}, where for a Weyl semimetal, one should swap $p_x \rightarrow p_\bot$. This produces a set of almost (up to a factor $\lambda$, see Eq.~\eqref{quasi-en0}, which is close to unity if $n$ is not very large) evenly-spaced Dirac points with identical spectrum. The Fermi velocity $v_{x\rm F}$ has a resonant behavior at field amplitudes
\vskip1mm
\begin{gather}
  \label{res1}
  U_{0k}=\frac{\pi\hbar (k-1/4) \omega_0}{2\lambda},
\end{gather}
%\vskip1mm
where $k$ is an integer. At these values of $U_{0k}$, the spectrum drastically changes. This is the manifestation of the resonant LZSM interferometry. At the values given by Eq.~\eqref{res1}, relation \eqref{quasi-en0} gives a flat band in the $x$ direction ($v_{x\rm F}$ vanishes at the resonance). This, of course, is the consequence of the quadratic expansion, which we used to find the spectrum. Retaining the last term in  phases~\eqref{phases1}, we obtain a very peculiar modification of the spectrum. Near the point $p_\bot=p_z=0$, the Floquet energy reads
\begin{gather}
  \label{res2}
 \begin{split}
   \ve_{\mathbf{p}}&=\pm\Bigg\{p_z^2+\frac{v^2p_\bot^2\omega_0}{\pi U_0}\Big(\sin^2\frac{p_\bot^2}{2U_0\omega_0}\ln\frac{8U_0}{\omega_0}\\
   &-\left(\frac{\pi v p_z}{\omega_0}\right)^2+\frac{4}{3}\left(\frac{\pi v p_z}{\omega_0}\right)^4\Bigg\}.
 \end{split}
\end{gather}
A numerical visualization at the resonant value corresponding to $n=4$ is presented in Fig.~\ref{graphene-spectrum}.
It is important to note that a spectrum similar to~\eqref{quasi-en0} was rederived in Ref.~\onlinecite{kibis1}. The method used in Ref.~\onlinecite{kibis1}, though allowing to lift the limit~\eqref{cond1}, completely fails to locate other non-trivial Dirac points at $p_{\bot}\neq0$. Taking into account just one trivial Dirac point is indeed possible, but in the case of a circularly polarized elecromagnetic wave of a small amplitude only. This case is addressed in Ref.~\onlinecite{kibis1}.

\section{Computation of the conductance\label{conduct}}
\subsection{Landauer--B\"{u}ttiker relation}
\begin{figure}[t] \centering
\includegraphics[width=0.9\columnwidth]{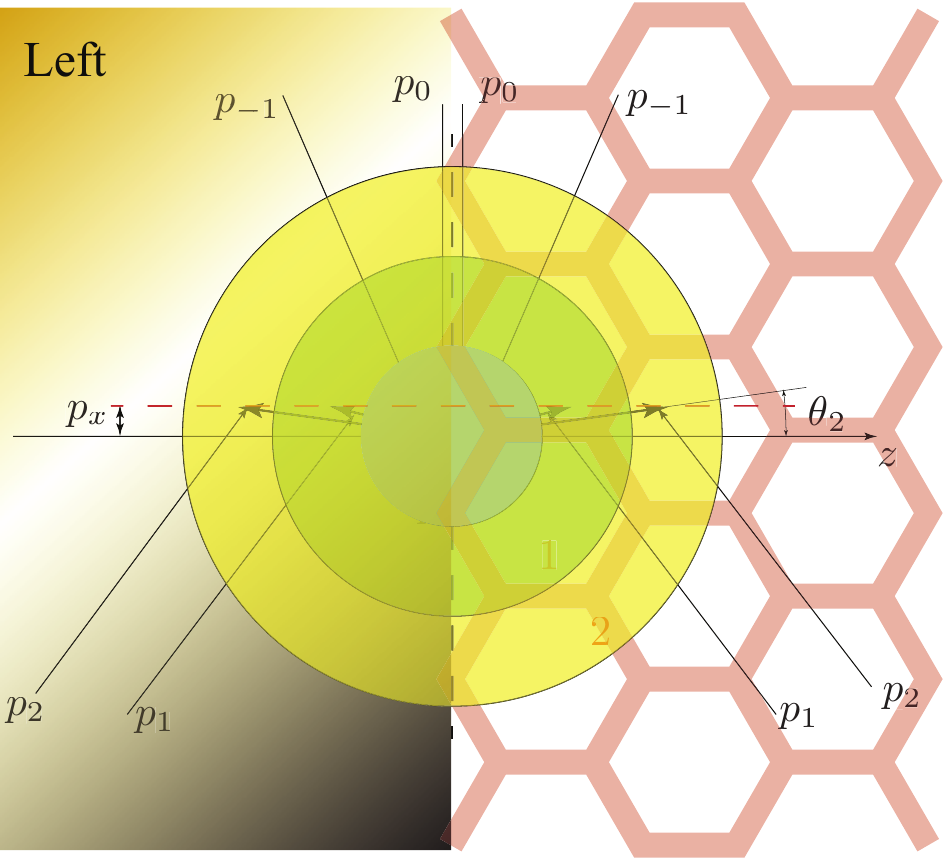}
\caption{(Color online) Floquet scattering. Different colors denote different Floquet zones. The momentum $p_0$ of the incident particle lies in the first Floquet zone $(n=0)$. $p_m=|\mu+\omega m|$.
     \label{floquet-scattering}
          }
\end{figure}

We now start the discussion of the driven conductance of the system. We are interested in the response of the electric current to the infinitesimal transport voltage $V$ (see Fig.~\ref{strip2}). The analytically solvable case corresponds to the limit of small chemical potential $\mu\ll\omega_0$.
It is also important to stress that it is the chemical potential of the contacts that is well-defined and enters all the equations.
The computation of a driven conductance is based on the corresponding non-stationary generalization of the Landauer--B\"{u}ttiker scattering matrix formula.~\cite{Landauer1957,Buttiker1985}
The schematics of the scattering process is presented in Fig.~\ref{floquet-scattering}. An incident charge carrier with energy $p_0$ undergoes scattering into Floquet bands picking up an integer number of energy quanta from the driving field: $p_0+\omega_0 m$. Although the system is non-stationary, the energy is still conserved modulo $\omega_0$ due to the periodicity of the external field. It is assumed that the contacts form a sharp edge between the irradiated and non-irradiated regions. The Floquet scattering takes place at both (left and right) edges of the contacts.
The zero-temperature conductance is then given by the following relation~\cite{SyzranovPRB2008}
\begin{gather}
  \label{landauer}
   G=e^2W\mu\int\frac{d\vf\sin\vf d\chi}{(2\pi)^3}\cos\vf\sum\limits_{m}
   |\mT_{m}(\ve)|^2\Big|_{\hskip-8mm\ve=\mu\atop \hskip-5mm p_{z,m}=\mu\cos\vf_m}.
\end{gather}
Here, $\mT_m(\ve)$ is the probability for an electron with energy $\ve$ in, say, the left reservoir  to traverse the irradiated region and get into the state $\ve+\omega_0 m$ in the right reservoir. $\vf_m$ is the angle at which the electron in the state $\ve+\omega_0 m$ moves in the right reservoir
(Fig.~\ref{floquet-scattering}). As one can see, only the amplitudes with initial energy $\ve=\mu$ contribute to the expression~\eqref{landauer}. In order to compute the conductance, we therefore need to solve the scattering problem.

\subsection{Solution of the Floquet scattering problem}

The wave functions inside the irradiated region were found in Ref.~\onlinecite{SRKN}.
We point out that the set-up in Fig.~\ref{strip2} preserves the translational invariance in the transverse (orthogonal to $z$) direction (we discard the influence of the boundaries, see the explanation in the Discussion section). Therefore, the momentum $p_\bot$ lying in the $\bot$ plane is conserved. Due to the conservation of energy modulo $\omega_0$, one deduces the new \textit{conservation} rule for the momentum in the longitudinal direction:
\begin{gather}
  \label{moment1}
   p_{z,m}=\pm\sqrt{(p_0+m\omega_0)^2-\bp_\bot^2},\ \ m\in Z.
\end{gather}

Now, we employ the condition
\begin{gather}
  \label{cond3}
  \frac{p_\bot}{\omega_0}\leq\beta\ll1.
\end{gather}
[see Eq.~\eqref{cond2}] The last inequality excludes the participation of the Dirac points at non-zero $p_\bot$ (see Fig.~\ref{graphene-spectrum}). Indeed, the Dirac points with non-zero $p_\bot$ are located at $p_\bot\sim n\omega_0,\ n\in Z$. This simplification plays a crucial role in the construction of an analytical solution. Even more, condition~\eqref{cond3} leads to the absence of evanescent modes in the scattering event ($p_{m,z}$ is real for $-\omega_0/2\leq p_0\leq\omega_0/2$).

The wave function within the irradiated and shaded regions is now sought in the form
\begin{gather}
     \Psi_\bp(t,\br)= \exp{(-i\ve_\bp t+i\bp_\bot\br_\bot)}\psi_\bp(t,z),
\end{gather}
where $\psi_\bp(t,z)$ is a periodic Floquet function.

The Floquet wave function $\psi_\bp(t,z)$ describing the scattering of the particles incident from the left lead is represented by the following relation
\begin{gather}
  \begin{split}
   &\psi_{\bp} (t,z)=\\
   &=
        \begin{cases}
          \Big[\exp{(ip_{z}z)}\psi(\ve_\bp,p_z)+ \\
          \sum\limits_{m} \mR_{m} \exp{[-i(\ve_{p_m} -\ve_\bp)t
          -i p_{m,z}z]}\psi(\ve_\bp,-p_{m,z})\Big],\quad z\leq 0\\
          \sum\limits_{m} \mT_{m} \exp{[-i(\ve_{p_m}-\ve_\bp)t+ip_{m,z}z]}\psi(\ve_\bp,p_{m,z}),\quad z\geq L,
        \end{cases}
  \end{split}
\end{gather}
where $\mT_m\equiv \mT(\ve,\ve+m\omega)$ and $\mR_m\equiv \mR(\ve,\ve+m\omega)$ are the transmission and reflection amplitudes, respectively, which relate the states
$\ve_\bp\rightarrow\ve_\bp+\omega_0 m$.
Due to conditions~\eqref{cond3}, the individual states $\psi_{m,z}$ entering the wave function have the form
\begin{gather}
   \label{states}
\begin{split}
   \psi(\ve_\bp,-p_{m,z})&\approx
  [1-\delta_{m,0}]\begin{pmatrix}
         0\\
         1
  \end{pmatrix}+\delta_{m,0}
  \begin{pmatrix}
         \sin\frac{\vf}{2}\\
         \cos\frac{\vf}{2}
  \end{pmatrix},\\
  \\
  \psi(\ve_\bp,p_{m,z})&\approx
  [1-\delta_{m,0}]\begin{pmatrix}
         1\\
         0
  \end{pmatrix}+\delta_{m,0}
  \begin{pmatrix}
         \cos\frac{\vf}{2}\\
         \sin\frac{\vf}{2}
  \end{pmatrix}.
  \end{split}
\end{gather}
Here we substituted
\begin{gather}
  \label{states2}
  \begin{split}
  \cos\frac{\vf_m}{2}&\approx 1+\mathcal{O}\left[\left(\frac{\mu}{\omega_0}\right)^2\right],\\
  \sin\frac{\vf_m}{2}&\approx \mathcal{O}\left(\frac{\mu}{\omega_0}\right)
  \end{split}
\end{gather}
and used condition~\eqref{cond2}.
%\subsection{The wave function inside the irradiated region}
As seen from Eqs.~\eqref{moment1}, the possible momenta of the scattered states $p_{z,m}=m\omega_0\pm\delta p_z,\ \ \delta p_z=\mu+\mathcal{O}(\mu^2/\omega_0)$ lie near the Dirac points.
Approximations \eqref{states} and \eqref{states2} allow for a considerable simplification of the scattering wave functions. Namely, the Floquet wave function takes the form
\begin{gather}
   \label{fl2}
  \begin{split}
   &\psi_{\bp} (t,z)=\\
   &=
        \begin{cases}
          \Big[e^{ip_{z}z}\begin{pmatrix}
                              \cos\frac{\vf}{2}&+&\sin\frac{\vf}{2}& \mR_0\\
                              \sin\frac{\vf}{2}&+&\cos\frac{\vf}{2}&[\mR_0-1]
                              +\mathcal{R}_t
                          \end{pmatrix}
          ,\quad z\leq 0
          \\
          \\
         \begin{pmatrix}
                              &\mT_0&[\cos\frac{\vf}{2}-1]&+&\mathcal{T}_t\\
                              &\mT_0&\sin\frac{\vf}{2}
                          \end{pmatrix},\quad z\geq L,
        \end{cases}
  \end{split}
\end{gather}
where we introduced the auxiliary scattering functions of time $\mR_t$ and $\mT_t$ defined as
\begin{gather}
  \begin{split}
   \mT_t&=\sum\limits_m \exp{(-im\omega_0 t)}\mT_m,\\
   \mR_t&=\sum\limits_m \exp{(-im\omega_0 t)}\mR_m.
  \end{split}
\end{gather}
The wave functions within the irradiated region are
\begin{gather}
   \psi_{\delta p_z}\approx
   \begin{pmatrix}
       &\exp{(-i\frac{U_0}{\omega_0}\cos t)}\\
       \\
        &\gamma \exp{(i\frac{U_0}{\omega_0}\cos t)}
   \end{pmatrix},\notag\\
   \\
      \psi_{-\delta p_z}\approx
   \begin{pmatrix}
       \gamma \exp{(-i\frac{U_0}{\omega_0}\cos t)}\\
       \\
          \exp{(i\frac{U_0}{\omega_0}\cos t)}
   \end{pmatrix}\, ,\notag\\
   \gamma=\frac{1}{2}v_{x\rm F}\sin\vf.
   \label{gamma-parameter}
\end{gather}
Here, only the terms of the order of $\sqrt{\kappa}$ are retained (note that in~\eqref{gamma-parameter}, $\gamma\sim\sqrt{\kappa}$) .

The sum in expression~\eqref{landauer} can now be simplified using the Fourier summation theorem
\begin{gather}
   \sum\limits_{m}|\mT_m|^2=\frac{1}{T}\int\limits_{0}^{T}\mathcal{T}^2_tdt
\end{gather}
Taking into account~\eqref{states2}, we obtain the following expression for the conductance
\begin{gather}
  \label{landauer1}
   \begin{split}
   &G=e^2W\mu^2\int\frac{d\vf\sin\vf}{(2\pi)^2\fv^3}\times\\
   &\times\left[\frac{1}{T}\int_0^{T}|\mathcal{T}_t(\vf)|^2dt
   +\mT_0^2(1-\cos\vf)\right]_{\ve(\bp)=\mu\atop p_z=\mu\cos\vf},\\
   \end{split}
\end{gather}
Here, the integration domain spans the  $\vf\in[0,\pi/2]$ range.

The Floquet wave function within the irradiated region can be represented by the following suitable parametrization
\begin{gather}
\begin{split}
 \label{state-graph}
  \psi_{\bp,G}&=
  \begin{pmatrix}
                    e^{-i\Theta_t}\left[f_{t-z}e^{i\delta p_z z}+\gamma g_{t-z}e^{-i\delta p_z z}\right]\\
                    e^{i\Theta_t}\left[f_{-t-z}e^{i\delta p_z z}+\gamma g_{-t-z}e^{-i\delta p_z z}\right]
  \end{pmatrix},\\
               \Theta_t&=i\frac{U_0}{\omega_0}\cos\omega_0t,
  \end{split}
\end{gather}
where $f_t$ and $g_t$ are arbitrary $T$-periodic functions (see Appendix for details). Here, one needs to understand the structure of the approximations made while computing the Floquet function within the irradiated region
\begin{gather}
   \psi_{G}=\alpha_1+\alpha_2\sqrt{\kappa}+\alpha_3\sqrt{\kappa}\beta
   +\alpha_4\kappa\beta+\alpha_5\kappa^{3/2}+...
\end{gather}
It is of utmost importance for us that the term proportional to $\kappa$ is absent in the expansion of the wave function in the region under study. The point is that the conductance has a quadratic dependence on the expansion parameters of the wave function $\psi_G$, $G\sim |\psi_G|^2$. The LZSM interferometry effects are hidden in  corrections of the order of $\kappa$. The condition~\eqref{cond2} allows us to omit the $\mu/\omega_0$ corrections to the wave functions inside the contacts as compared to the terms $\sim\kappa$.

The general structure of scattered states~\eqref{states} gives us a clear pattern of the scattering event. The scattered states with energy $\ve_\bp+m\omega_0,\ m\neq0$ move almost orthogonally to the interface due to the smallness of the transverse velocity $v_{x\rm F}$. As a result, the functions of the scattering angle can be written as $\sin\vf_n=\mathcal{O}[(\mu/\omega_0)],\ \ \cos\vf_n=1+\mathcal{O}[(\mu/\omega_0)^2]$.
Therefore, we take $\sin\vf_n\approx0,\ \cos\vf_m\approx1$ when solving the system of equations. Next, one should match the solutions at $z=0$ and $z=L$.

As a result, we arrive at the following set of equations
\begin{gather}
   \begin{split}
      f_t+\gamma g_t&=e^{-i\Theta_t}\left[\cos\frac{\vf}{2}+r_0\sin\frac{\vf}{2}\right],\\
      \gamma f_t+g_t &=e^{i\Theta_t}\left[R_t+\sin\frac{\theta}{2}+r_0(\cos\frac{\vf}{2}-1)\right],\\
   \end{split}
\end{gather}
\begin{gather}
   \begin{split}
      e^{-i\Theta_t}\left[f_{t-L}e^{i\delta p_z L}+\gamma g_{t-L} e^{-i\delta p_z L}\right]&=t_0\cos\frac{\vf}{2}-t_0+
      \mT_t,\\
      e^{i\Theta_{t+L}}\left[g_te^{-i\delta p_z L}+\gamma f_t e^{i\delta p_z L}\right]&=t_0\sin\frac{\vf}{2} .
  \end{split}
\end{gather}
After simple but cumbersome algebra, one obtains  the following solution for the scattering function up to the second order in $\sqrt{\kappa}$
\begin{gather}
 \label{amplitude}
  \begin{split}
   &\mathcal{T}_t=\frac{\cos\vf}{\cos\frac{\vf}{2}}e^{-\theta_t+i\theta_{t-L}+i\delta p_z L}\left(\frac{\cos\vf}{\cos\frac{\vf}{2}}+
  \delta \mR_0^{(2)}\sin\frac{\vf}{2}\right)\\
  &+t_0^{(1)}\left(1-\cos\frac{\vf}{2}\right)-2i\gamma\sin(\delta p_z L)\bigg[\mT_0^{(1)}e^{-2i\theta_{t}}\sin\frac{\vf}{2}\\
  &-2i\gamma\sin(\delta p_z L)e^{i\theta_{t-L}-i\theta_t}\frac{\cos\vf}{\cos\frac{\vf}{2}}\bigg]e^{i\delta p_z L}\\
  &+2i\sin(\delta p_z L)\gamma^2e^{i\theta_{t-L}-i\theta_t}\frac{\cos\vf}{\cos\frac{\vf}{2}},
  \end{split}
\end{gather}
where
\begin{gather}
 \label{tr}
  \begin{split}
   \mT_0^{(1)}&=\frac{\cos\vf}{\cos^2\frac{\vf}{2}}e^{i\delta p L}J_0\left(\frac{2U_0}{\omega_0}\sin\frac{\omega_0 L}{2}\right)\\
   \delta \mR_0^{(2)}&=\frac{e^{i\delta p L}}{\cos\frac{\vf}{2}}\bigg[\mT_0^{(1)}\sin\frac{\vf}{2} J_0\left(\frac{2U_0}{\omega_0}\sin\frac{\omega_0 L}{2}\right)\\
   &-2i\gamma\sin\delta p_z L J_0\left(\frac{2U_0}{\omega_0}\right)\frac{\cos\vf}{\cos\frac{\vf}{2}}\bigg].
  \end{split}
\end{gather}
Next, we substitute expressions \eqref{amplitude} and \eqref{tr} into  Eq.~\eqref{landauer1}.

\begin{figure}[h]
\includegraphics[width=0.9\columnwidth]{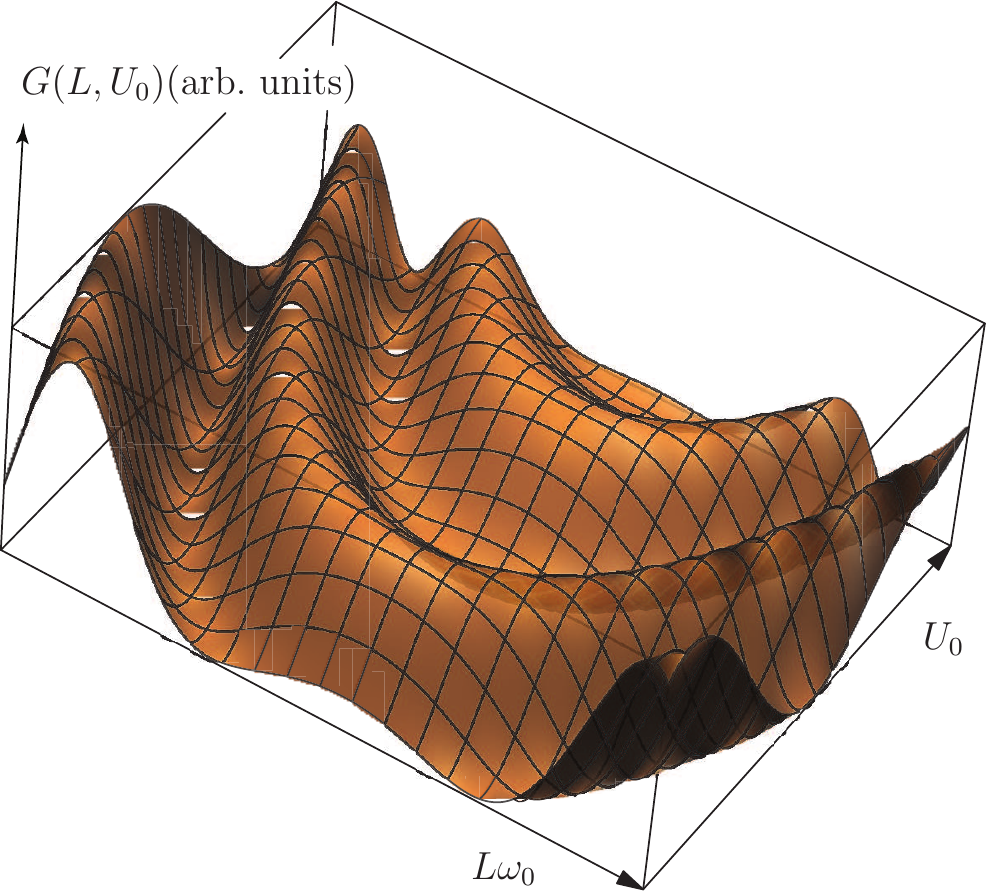}
\caption{\label{conduct1}(Color online)
Driven conductance vs external field amplitude $U_0$ and the length of the sample $L$ calculated according to Eq.~\eqref{conduct-final}. The oscillations as a function of $L$ and $U_0$ are clearly pronounced. The main contribution to the amplitude comes from the Ramsauer--Townsend effect. More subtle LZSM oscillations are illustrated in Fig.~\ref{splot}.
          }
\end{figure}

\begin{figure}[t] \centering
\includegraphics[width=0.9\columnwidth]{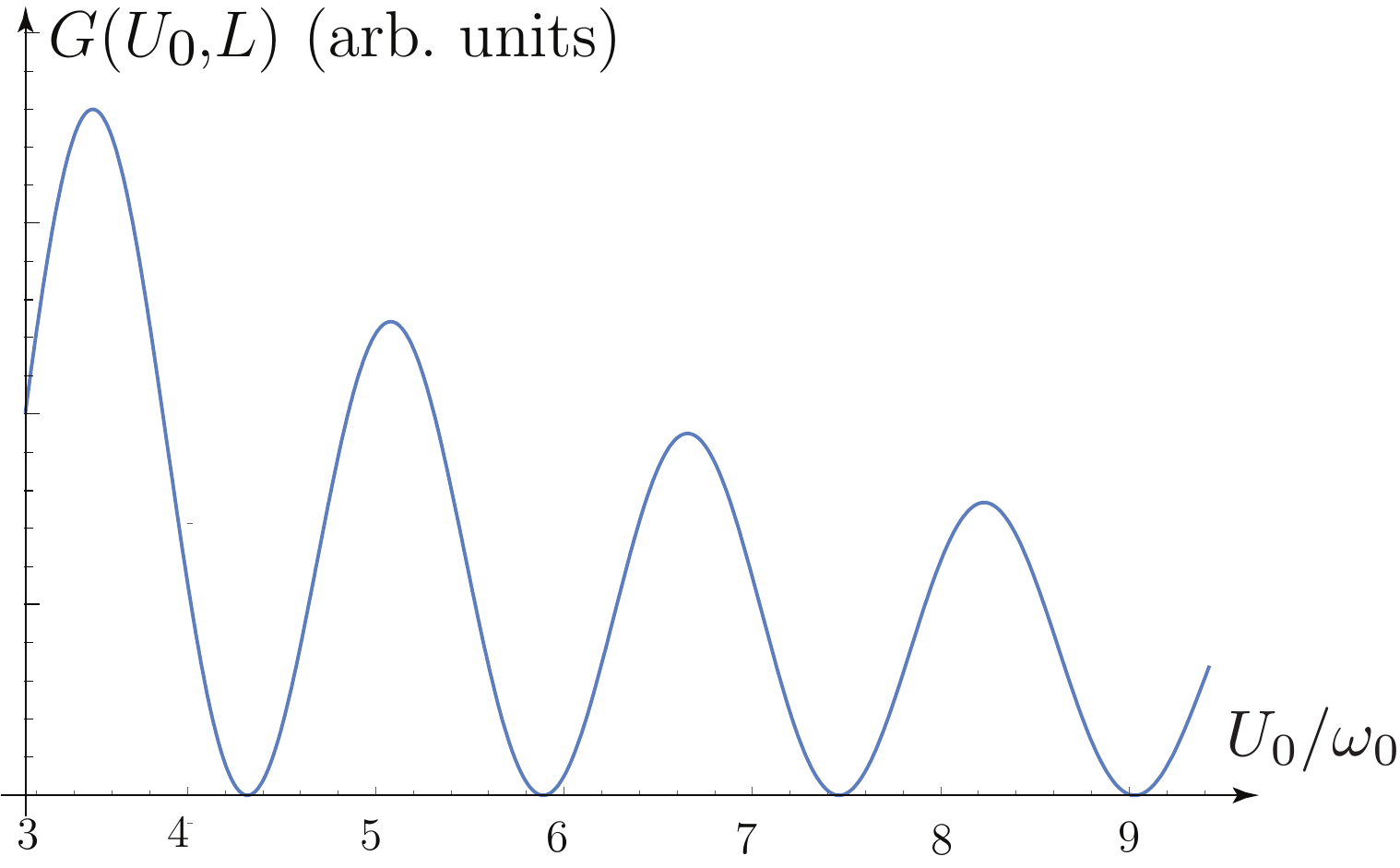}
\caption{(Color online) Conductance oscillations proportional to the strongly renormalized velocity $v_x(U_0)$ of the Floquet excitations. The velocity renormalization is the manifestation of LZSM interferometry. Here, $\mu=0.05\omega_0$.
     \label{splot}
          }
\end{figure}
Retaining the lowest-order terms, we obtain the final formula for the driven conductance
\begin{gather}
\label{conduct-final}
 \begin{split}
    &G(\omega_0,U_0)=\frac{e^2W\mu^{D-1}\beta_0}{2\pi\fv^D}\\
    &\times\bigg[
       1+\beta_1 J_0^2\left(\frac{2\fv U_0}{\hbar\omega_0}\sin\frac{\omega_0 L}{2\fv}\right)(2\cos\frac{2\mu L}{\hbar\fv}+1)\\
    &+\beta_2 \frac{v^2_{x\rm F}}{\fv^2}\sin^2\frac{\mu L}{\hbar\fv}
    \bigg],
  \end{split}
    \end{gather}
where
\begin{gather}
 \begin{split}
   \beta^{3D}_0&=\frac{\frac{5}{3}-\ln 4}{2\pi},\ \  \beta^{3D}_1=\frac{14\ln 2\!-\!\frac{29}{3}}{\frac{5}{3}-\ln 4},\ \ \beta_2^{3D}=\frac{\frac{167}{40}\!-\!4\ln 4}{\frac{5}{3}-\ln 4},\\
   \beta^{2D}_0&=\frac{3\pi}{2}-4,\ \  \beta^{2D}_1=\frac{\frac{104}{9\pi}-\frac{11}{3}}{1-\frac{8}{3\pi}},\ \ \beta_2^{2D}=\frac{\frac{17}{4}-\frac{40}{3\pi}}{1-\frac{8}{3\pi}}.
   \end{split}
\end{gather}
Here, $D=3$ for Weyl semimetal and $D=2$ for graphene, $\beta^{3D,(2D)}$ correspond to a  Weyl semimetal or graphene, respectively and $v_{x\rm F}$ is given by \eqref{quasi-en0}.

\section{Discussion and conclusions}
The analytical expression ~\eqref{conduct-final} describing the conductance under the effect of a strong electromagnetic wave is the central result of our paper.
It takes into account the contribution of \textit{all} Dirac points existing in the spectrum of Floquet excitations. As we will see, in the most experimentally viable situations, the argument $\mu L\ll1$ and  $\cos,\sin(\mu L)$ do not oscillate. The important oscillating terms are those, which contain the semiclassical large pre-factors $U_0/\omega_0$. These are  $J_0^2$ and $v_{x,\rm}^2$ terms in~\eqref{conduct-final}.

One immediately notices that the driven conductance exhibits oscillations as a function of the external field amplitude $U_0$. There are two oscillatory contributions. The first one has the following specific feature. It has an oscillatory dependence on the external field $U_0$ \textit{as well as} on the length of the exposed area $L$. This (related to $\omega$) $L$-dependence is a manifestation of the Ramsauer--Tounsend effect.~\cite{Griffits}

The second oscillating term is the most interesting one. It is proportional to the velocity squared $v_{x\rm F}^2$ of the Floquet modes in the $x$ direction. It is the manifestation of the presence of quasienergy excitations in the irradiated area.
%his term is therefore the manifestation of the presence of Floquet excitations and, as a result, the LZSM-physics.
From the experimentalist's point of view, it is easier to measure the conductance as a function of the external field intensity and frequency. To get rid of the Ramsauer--Tounsend oscillations, one can tune the frequency of the external field in such a way that  $J_0^2$ in the second term of~\eqref{conduct-final} disappears [see Eq.~\eqref{omega}].

While solving the system of scattering equations, we completely discarded the influence of the parameter $\lambda$. We took it approximately to be $\lambda\sim1$. We, therefore, assumed that the major contribution to the solution of the scattering problem comes from the Fourier components of $\mR_n$ and $\mT_n$ with not very large numbers: $n\lesssim U/\omega_0$. To check this approximation we developed a numerical scheme for the computation of the conductance.

The numerics were performed by an exact calculation of the evolution operator
\begin{gather}
   U(t)=T \exp{\left(-i\int_0^t H(t)\,dt \right)},
\end{gather}
on a time grid with the length equal to the period of the external field and a slice $2\pi/(N\omega_0)$. The eigenvalues of this operator determine the Floquet functions. We took several values of $N$ ($N=$ 80, 90, and 100) to check the stability of the numerical scheme. $(N-4)/2$ Dirac points of Floquet states were taken into account to form a closed linear system to solve the scattering problem. The numerics and theory slightly deviate from each other. In Fig.~\ref{conduct1}, we present the calculated conductance surfaces as a function of $U_0$ and $L\omega$. The numerical and theoretical plots turn out to be nearly indistinguishable. %The surface shape provides, of course, just a qualitative comparison.

To illustrate the contribution from the Floquet excitations, we show in Fig.~\ref{splot} the behavior of the third term in  Eq.~\eqref{conduct-final} as a function of $U_0$. The behavior obtained numerically is qualitatively the same, but differs in the amplitude of the oscillations. To get rid of the discrepancy, one needs to treat the $\mR_n$ and $\mT_n$ modes with higher accuracy. This significantly complicates the analytical approach.  We leave this for future work.

Now we need to check the experimental viability of the obtained theoretical conductance. We completely omitted the influence of the disorder. The elastic scattering time was estimated in recent experiments. Experiment~\cite{Liang2015} explored ${\rm Cd}_3{\rm As}_2$ and obtained $\tau_{\rm el}\sim 10^{-13}\, s$. Experiment~\cite{Zhang2016} gives $\tau_{\rm el}\sim 10^{-12}\, s$ in TaAs. That provides the lowest limit for a possible radiation field frequency $\omega\ge10^{12}$ Hz in ${\rm Cd}_3{\rm As}_2$ and $\omega\ge 10^{13}$ Hz in TaAs. The chemical potential should be $\mu\lesssim 1$ meV (in ${\rm Cd}_3{\rm As}_2$) or $\mu\lesssim 10$ meV (in TaAs). The typical length of the sample $L\sim 1\,\mu{\rm m}$ corresponds to a $L\mu$ factor of the order of $\lesssim 1$.
The experimental value of the chemical potential in Cd$_3$As$_2$ is about 50 meV.\cite{giant2} As we see, it is much larger than 1 eV, allowed by the derivation. However, as we mentioned in the text earlier, $\mu$ is the chemical potential of electrons injected into the contacts. Recent experiment \cite{gate} shows that the chemical potential of a thin layer of WSM can be controlled by the gate voltage (even through the Weyl point) in a perfect analogy to graphene. Note, that according to Refs. \onlinecite{Skinner} and \onlinecite{RS}, the Weyl semimetal should not be highly compensated to prevent the formation of electron puddles.

To conclude, we studied the driven conductance of a Dirac material (either 2D or a thin 3D film) in strong linearly-polarized electromagnetic field. We discovered that the driven conductance is the observable that allows one to see the manifestation of Floquet physics hidden in the irradiated region of a semimetal. The LZSM interferometry is responsible for strong oscillations of the renormalized Fermi velocity of the Floquet excitation and this is precisely the quantity, which causes the oscillatory behavior of the driven conductance.

\section*{Acknowledgements}
%%%%%%%%%%%%%%%%%%%%%%%%%%%%%%%%%%%%%%%%%%%
\label{ackno}
We are grateful to S. Syzranov, S. Shevchenko, O. Kibis, and Zhou Li for critical reading the manuscript and useful remarks.
This work is partially supported by the Russian Foundation for Basic Research (projects 14-02-00276 and 15-02-02128), RIKEN iTHES Project, the MURI Center for Dynamic Magneto-Optics via the AFOSR award number FA9550-14-1-0040, the IMPACT program of JST, a Grant-in-Aid for Scientific Research (A), and a grant from the John Templeton Foundation. YaIR gratefully acknowledges the financial support of the Ministry of Education and Science of the Russian Federation in the framework of the Increase Competitiveness Program of NUST MISIS (grant K2-2015-076).

\appendix*
\section{LZSM interferometry}
Here, we present the details of the calculations of the LZSM transfer matrices. Since the phases of the two consequential LZSM amplitudes play a crucial role, we feel it necessary to rederive all the amplitudes to be on the safe side. We hope that the details of the derivation will be of some use for solving other related problems. The method involved can be found in books and papers dealing with the asymptotic analysis (see e.g. Ref.~\onlinecite{Aoki}).

We solve the original system of differential equations on eigenfunctions
\begin{gather*}
    i\frac{\de\psi}{\de t}=H\psi
\end{gather*}
as
\begin{gather*}
   \ddot{\psi}_\uparrow-i\dot{U}\psi_\uparrow+[(U-p_z)^2+p_\bot^2]\psi_\uparrow=0,\\
   \psid=\frac{i\hbar\dot{\psi}_\uparrow+(U-p_z)\psiu}{p_\bot}.
\end{gather*}
If the semiclassical condition
\begin{gather}
  \label{lzsm1}
p_\bot\ll U_0
\end{gather}
holds, then at points defined by the equation $U_0\sin\omega t_{1,2}=p_z$, the LZSM transitions take place
\begin{gather}
  \begin{split}
   \omega t_1=\arcsin\frac{p_z}{U_0},\quad
      \omega t_2=\pi-\arcsin\frac{p_z}{U_0}.
  \end{split}
\end{gather}
We denote $u_1=\dot{U}|_{t=t_1}=-\dot{U}|_{t=t_2}$.
Even more, away from the transition points, the wave function obeys a semiclassical evolution and is given by Eq.~\eqref{gen-sol}.

Next, we expand the potential $U(t)$ near the crossing points $t_{1,2}$. The expansion is legitimate if and only if condition~\eqref{lzsm1} is satisfied.
Making a suitable change $s=\sqrt{u_1}(t-t_{1,2})$, we arrive at two equations, describing the system near the corresponding LZSM transition:
\begin{gather}
  \label{lz1}
 \begin{split}
  \ddot{\psi}+[s^2+\nu-i]\psi&=0\ \ \hbox{first LZSM transtion},\\
  \ddot{\psi}+[s^2+\nu+i]\psi&=0\ \ \hbox{second LZSM transtion}.
 \end{split}
\end{gather}
In what follows, we perform a full derivation of the LZSM transfer matrix for both transitions. First, we present the semiclassical expressions in accordance to~\eqref{gen-sol}. Expanding the \textit{semiclassical momentum} $q=\sqrt{(U(t)-p_z)^2+p_\bot^2}=u_1\sqrt{s^2+\nu}$ near the transition points $t_1$ and $t_2$ and performing simple integrals and algebra we arrive at
\begin{eqnarray}
   \label{asy-gtr1}
   \psi^{(1)}_{\underset{{\rm semi}+}{s>0\uparrow}}&=
   \Big(\frac{2s}{\sqrt{\nu}}\Big)^{\frac{i\nu}{2}}
   \exp\left(\frac{is^2}{2}+\frac{i\nu}{4}\right),\\
%%%%%%%%%%%%%%%%%%%%%%%%%%%%%%%%%%%%%%%%%%%%%%%%%%%%%%%%%%%%%%%%%%%%%%%%
     \label{asy-less1}
    \psi^{(1)}_{\underset{{\rm semi}+}{s<0\uparrow}}&=
   \Big(\frac{-2s}{\sqrt{\nu}}\Big)^{\frac{i\nu}{2}}
   \exp\left(\frac{is^2}{2}+\frac{i\nu}{4}\right).
\end{eqnarray}
\begin{eqnarray}
   \label{asy-gtr2}
   \psi^{(1)}_{\underset{{\rm semi}-}{s>0\uparrow}}&=
   \Big(\frac{2s}{\sqrt{\nu}}\Big)^{-\frac{i\nu}{2}-1}\exp\left(-\frac{is^2}{2}
   -\frac{i\nu}{4}\right),\\
%%%%%%%%%%%%%%%%%%%%%%%%%%%%%%%%%%%%%%%%%%%%%%%%%%%%%%%%%%%%%%%%%%%%%%%%%%
   \psi^{(1)}_{\underset{{\rm semi}-}{s<0\uparrow}}&=-
   \Big(\frac{-2s}{\sqrt{\nu}}\Big)^{-\frac{i\nu}{2}-1}
   \exp\left(-\frac{is^2}{2}-\frac{i\nu}{4}\right).\quad\quad\ \
\end{eqnarray}
\begin{eqnarray}
   \label{asy-gtr3}
   \psi^{(2)}_{\underset{{\rm semi}+}{s>0\uparrow}}&=
   \Big(\frac{2s}{\sqrt{\nu}}\Big)^{-\frac{i\nu}{2}}
   \exp\left(-\frac{is^2}{2}-\frac{i\nu}{4}\right),\\
%%%%%%%%%%%%%%%%%%%%%%%%%%%%%%%%%%%%%%%%%%%%%%%%%%%%%%%%%%%%   \label{asy-less3}
   \psi^{(2)}_{\underset{{\rm semi}+}{s<0\uparrow}}&=
   \Big(\frac{-2s}{\sqrt{\nu}}\Big)^{-\frac{i\nu}{2}}
   \exp\left(-\frac{is^2}{2}-\frac{i\nu}{4}\right).\quad\quad\ \
\end{eqnarray}
\begin{eqnarray}
   \label{asy-gtr4}
   \psi^{(2)}_{\underset{{\rm semi}-}{s>0\uparrow}}&=
   -\Big(\frac{2s}{\sqrt{\nu}}\Big)^{\frac{i\nu}{2}-1}
   \exp\left(\frac{is^2}{2}+\frac{i\nu}{4}\right),\\
%%%%%%%%%%%%%%%%%%%%%%%%%%%%%%%%%%%%%%%%%%%%%%%%%%%%%%%%%%%%
   \label{asy-less4}
   \psi^{(2)}_{\underset{{\rm semi}-}{s<0\uparrow}}&=
   \Big(\frac{-2s}{\sqrt{\nu}}\Big)^{\frac{i\nu}{2}-1}
   \exp\left(\frac{is^2}{2}+\frac{i\nu}{4}\right).\quad\quad\ \
\end{eqnarray}
Now, we build formal exact solutions of equations~\eqref{lz1}. First, we turn them into equations with linear coefficients by using the substitution
\begin{gather}
  \label{change1}
   \psi=\exp\left(\pm is^2/2\right)\vf(s).
\end{gather}
We obtain
\begin{gather}
  \label{lz2}
 \begin{split}
     \ddot{\vf}+2si\dot{\vf}+\nu\vf&=0,\\
    \ddot{\vf}-2si\dot{\vf}+\nu\vf&=0.
 \end{split}
\end{gather}
Then, we use the standard Laplace technique to write down the solutions in the form of complex integrals
\begin{eqnarray}
  \label{exact1}
   \vf(s)&=\int\limits_{C_{1,2}} \exp\left(st-it^2/4\right)t^{-i\nu/2-1}\,dt,\\
    \label{exact2}
   \vf(s)&=\int\limits_{C_{3,4}} \exp\left(st+it^2/4\right)t^{i\nu/2-1}\,dt.
\end{eqnarray}
The position and the shape of the contours is defined by the condition that the function
\begin{gather}
 \label{v-func}
  V_{1,2}=t^{-i\nu/2}\exp\left(st\mp\frac{it^2}{4}\right)
\end{gather}

\begin{widetext}
\begin{figure}[H] \centering
%\captionsetup{width=18cm}
\includegraphics[width=0.9\textwidth]{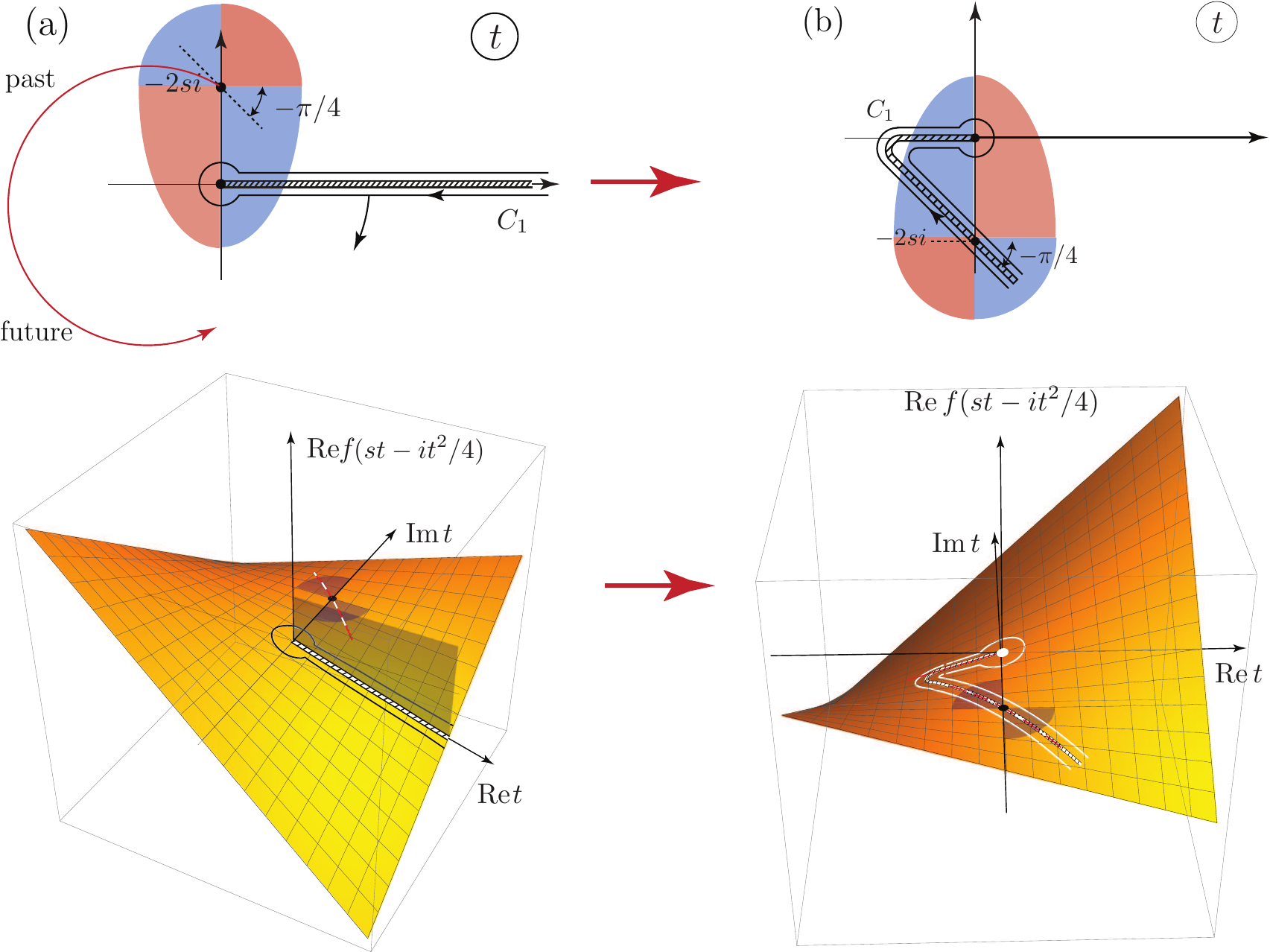}
\caption{(Color online) The contour defining the asymptotics of the solution~\eqref{exact1} in the complex plane of the $t$ variable for the first LZSM transition.  Blue areas mark the regions where $\hbox{Re}[st-it^2/4]$ is smaller than the value of the same function at the saddle point. These are the \textit{allowed} regions for the contour of integration. Below, the relief of the real part of the exponential function ${\rm Re}\,(st-it^2/4)$ is presented.
     \label{contour-ss1a}
          }
\end{figure}
\end{widetext}
has identical values at the end points of a given contour.
The integrands in \eqref{exact1} and \eqref{exact2} as well as the functions~\eqref{v-func} are multivalued. The main problem is how to draw branch cuts and contours $C_{i}$ in such a way that solutions~\eqref{exact1}, \eqref{exact2} would yield the correct asymptotics. We explain how it is done with solution~\eqref{exact1}. All other asymptotics are obtained in a similar manner.

First, let us look at the semiclassical solution~\eqref{asy-less1}. It is the wave coming from the \textit{infinite past} $s\rightarrow-\infty$. It is also going to be the asymptotics of an exact solution~\eqref{exact1}.

\subsection{First transition: solution $\sim \exp(is^2/2),\ s\rightarrow-\infty$}

We characterize this solution by its dominant exponential term $\exp(is^2/2)$. After the LZSM transition, when the time tends to infinite future, $s\rightarrow+\infty$, the asymptotics of the solution will pick up the part of the other linear independent solution~\eqref{asy-gtr2} (apart from ~\eqref{asy-gtr1}, which is a simple analytical continuation of the solution~\eqref{asy-less1}).

Now, we analyze the integrand of~\eqref{exact1}.  At $|s|\gg1$. the behavior of the integral is governed by the exponent $st-it^2/4$. It has a saddle point at $t=-2si$. The steepest descent path is inclined at an angle $3\pi/4$ with respect to the real axis of $t$ [see the relief of ${\rm Re}(st-it^2/4)$ in Fig.~\ref{contour-ss1a}]. The contribution from this saddle leads to the $\psi\sim \exp(-is^2/2)$ term at $s\rightarrow-\infty$. Obviously, this is not what we need. Therefore, the position of the branch cut should be chosen in a way prohibiting the contour to pass through the saddle point. Thus, it is clear that the branch cut needs to go to infinity in the right complex semiplane.

Next, we discuss condition~\eqref{v-func}. The regions where $V(t,s)$ decays (tends to zero) are the second and fourth quadrants of the complex plane, i.e. they approximately coincide with the \textit{allowed} (blue) regions defined by the saddle point (see Fig.~\ref{contour-ss1a}(a) and explanation to it).  The only way to draw a contour, which yields a nonzero solution and cannot be deformed to pass through the saddle point, is the path, which fully encircles the branch cut (under condition that the branch cut itself has the ending in the second or fourth quadrant). The exact direction of the branch cut still has not yet been fixed. It is determined by the following argument. The placement of the branch cut should facilitate the extraction of the asymptotics. If the contour does not traverse the saddle point, the asymptotics is defined by the vicinity of the branch point $t=0$ and the subsequent integration along the steepest descent direction.
For $s<0$ the steepest descent from point $t=0$ is the positive ${\rm Re}\,t$. This fixes the placement of the branch cut, see Fig.~\ref{contour-ss1a}(a), contour $C_1$. The asymptotics is obtained trivially by dropping the term $it^2/4$ in the exponential
\begin{gather}
  \label{func-asy1}
  \vf(s)=\Gamma\left(-\frac{i\nu}{2}\right)\left(1-e^{\pi\nu}\right)(-s)^{i\nu/2},\ \ -s\gg1.
\end{gather}

One arrives (in combination with the prefactor $\exp(is^2/2)$)  at the desired behavior at $s\rightarrow-\infty$. When we traverse to the $s>0$ region, the relief of function ${\rm Re}(st-it^2/4)$ is changed, entailing the change of the topology of an integration path. Suppose that we travel from the infinite past to the infinite future  (from large negative $s$ to large positive $s$) via the rotation in the lower half-plane (counterclockwise, $\Delta\hbox{arg}\, s=\pi$) of the time domain. How shall we deform the contour to compute the asymptotics?

The peculiar thing is that the steepest descent direction at the branch point is rotated clockwise simultaneously to preserve the negative sign of the $st$ factor in the exponential in~\eqref{exact1}. On the other hand, the saddle point itself moves in the complex plane [see the lower part of Fig.~\ref{contour-ss1a}(a)]. The end of the contour must not leave the specified allowed \textit{blue} region of the complex plane.

Therefore, we have to bend the contour. However, only the lower part of the contour can follow the steepest descent path (clockwise) from the branch point. The path of the upper part of the contour is blocked by the branch cut. This means that the upper part of the contour slips onto the upper sheet of the Riemann surface of the multivalued function $t^{i\nu/2}$. One can avoid this detour into 3D by the following trick. Instead of going on to the other Riemann sheet, one can \textit{bend the branch cut} in such a way that it does not block the path of the upper part of the contour, see Fig.~\ref{contour-ss1a}(b)). One can then further force the branch cut (and therefore, the contour) to pass through the saddle point. This deformation solves the problem of extricating the asymptotics at $s\rightarrow+\infty$.

As a result, there is one more contribution to the asymptotics coming from the saddle point. This is precisely the contribution that describes the LZSM transition.

The saddle point is passed in two directions: to the left of the branch cut and to the right. Therefore, we can identify the argument of $t$ as $3\pi/2$ to the left and $-\pi/2$ to the right. The contribution from a saddle point is
\begin{gather}
    \sqrt{4\pi}(2s)^{-i\nu/2-1}e^{i\pi/4}e^{-\pi\nu/4}(1-e^{\pi\nu}).
\end{gather}
Hence, the total asymptotics reads
\begin{gather}
   \label{func-asy2}
   \vf(s)=-2\sinh\frac{\pi\nu}{2}\Gamma\left(-\frac{i\nu}{2}\right)s^{i\nu/2}\\
   +\sqrt{4\pi}(2s)^{-i\nu/2-1}e^{i\pi/4}e^{-\pi\nu/4}(1-e^{\pi\nu})\exp(-is^2),\ \ s\gg1.\notag
\end{gather}
Collecting~\eqref{func-asy1} and~\eqref{func-asy2}
and dividing by $\Gamma(-i\nu/2)(1-e^{\pi\nu})$, we obtain the correct asymptotic behavior of function $\vf(s)$ in the form
\begin{gather}
   \vf(s)=\begin{cases}
              (-s)^{i\nu/2},\ \ -s\gg1,\\
              \, \\
              e^{-\pi\nu/2}s^{i\nu/2}
              +\sqrt{\nu}\sqrt{1-e^{-\pi\nu}}e^{i{\rm arg}\,\Gamma+i\pi/4},\\
              \times e^{-is^2}(2s)^{-i\nu/2-1},\ \ s\gg1.
          \end{cases}
\end{gather}
Here, we used the identity:
\begin{gather}
  \left|\Gamma\left(ix\right)\right|=\sqrt{\frac{\pi}{x\sinh\pi x}}.
\end{gather}
\begin{figure}[tbh] \centering
\includegraphics[width=0.8\columnwidth]{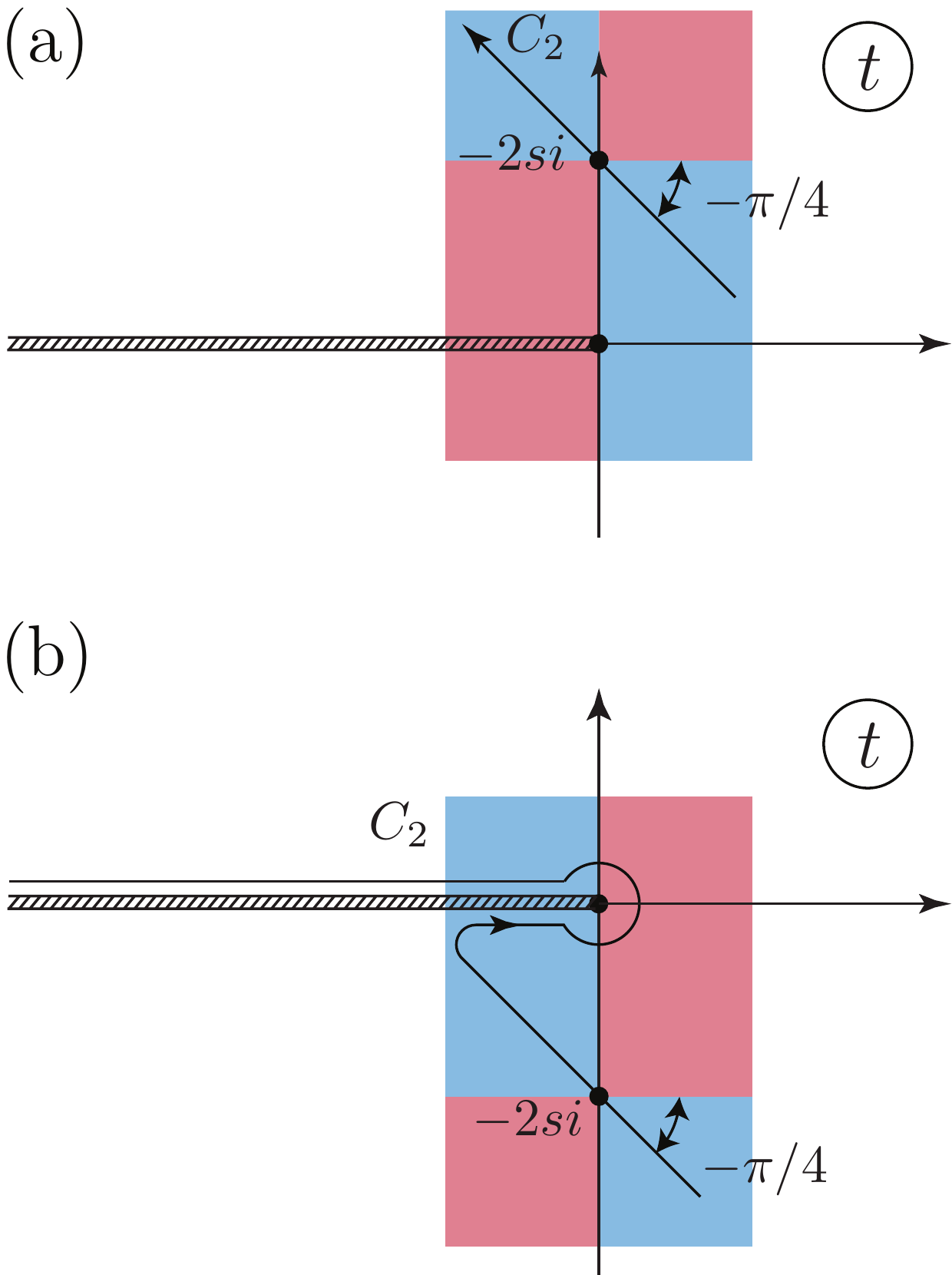}
\caption{(Color online) Contours defining the asymptotics of the solutions in the complex plane of the $t$ variable for the first LZSM transition.
     \label{contour-s1b}
          }
\end{figure}

The contours defining the linearly independent solutions and the deformations defining the asymptotics are shown in Fig.~\ref{contour-s1b} and~\ref{contour-s2}, respectively.
We also present the correct contour deformation, which gives the asymptotics.

Finally, we present results for all other asymptotics.

\subsection{First transition: solution $\sim\exp(-is^2/2),\ s\rightarrow-\infty$}

\begin{figure}[H] \centering
\includegraphics[width=0.8\columnwidth]{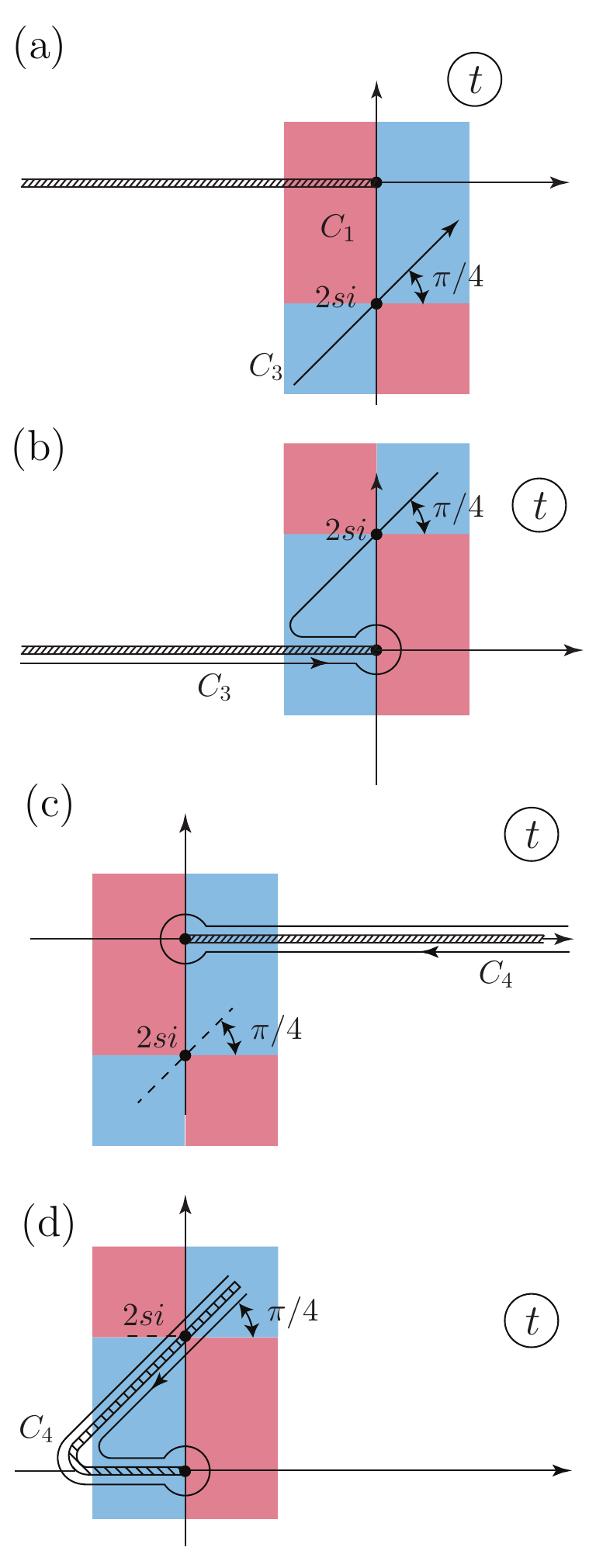}
\caption{(Color online) Contours defining the asymptotics of the solutions in the complex plane of the $t$ variable for the second LZSM transition.
     \label{contour-s2}
          }
\end{figure}

The asymptotics at $s\rightarrow-\infty$ is given by
\begin{gather}
  \label{asy-ex1}
   \vf(s)=\sqrt{4\pi}(-2s)^{-i\nu/2-1}e^{i\pi/4}e^{\pi\nu/4}\exp(-is^2),\ \ s\rightarrow-\infty.
\end{gather}
When $s\rightarrow+\infty$, the saddle point changes its position and the contour is deformed in the way shown in Fig.~\ref{contour-s2}(b).
Then, there are two contributions: from the saddle point and from the vicinity of a branch point.
Hence
\begin{gather}
   \vf(s)=\sqrt{4\pi}(2s)^{-i\nu/2-1}e^{3\pi i/4}\left(e^{-i\pi/2}\right)^{-i\nu/2-1}\exp(-is^2)\notag\\
   +s^{i\nu/2}2\sinh\frac{\pi\nu}{2}\Gamma\left(-\frac{i\nu}{2}\right).
   \label{asy-ex2}
\end{gather}
Matching \eqref{asy-ex1} and \eqref{asy-ex2} with the semiclassical expressions, and combining asymptotics $\sim \exp(\pm is^2/2)$ at $s\rightarrow-\infty$ , we obtain the LZSM transfer matrix for the first transition (eq.~\eqref{lz-matrix} upper signs).

%%%%%%%%%%%%%%%%%%%%%%%%%%%%%%%%%%%%%%%%

\subsection{Second transition: solution $\sim\exp(is^2/2),\ s\rightarrow-\infty$}

See Fig.~\ref{contour-s2} for the correct placement of branch cuts and deformation of the contours.
The asymptotics when $s<0$ is given by the saddle point
\begin{gather*}
   \vf(s)=\sqrt{4\pi}(-2s)^{i\nu/2-1}\left(e^{-i\pi/2}
   \right)^{i\nu/2-1}e^{is^2}e^{i\pi/2}\\
   =\sqrt{4\pi}e^{\pi\nu/4+3\pi i/4}e^{is^2}(-2s)^{i\nu/2-1},\ \ s\rightarrow-\infty .
\end{gather*}

When $s>0$, we have
\begin{gather*}
  \vf(s)=\sqrt{4\pi}(2s)^{i\nu/2-1}e^{is^2-\pi\nu/4-i\pi/4}\\
  -2\sinh\frac{\pi\nu}{2}\Gamma\left(\frac{i\nu}{2}\right)s^{-i\nu/2},\ \ s\rightarrow+\infty .
\end{gather*}

\subsection{Second transition: solution $\sim\exp(-is^2/2),\ s\rightarrow-\infty$}
For $s<0$, the main contribution comes from the vicinity of $t=0$
\begin{gather*}
  \vf(s)=\Gamma\left(\frac{i\nu}{2}\right)\frac{1-e^{-\pi\nu}}{(-s)^{i\nu/2}}.
\end{gather*}
For $s>0$, there is an additional contribution coming from the saddle point (two paths on each bank of a branch cut)
\begin{gather*}
  \vf(s)=\Gamma\left(\frac{i\nu}{2}\right)(1-e^{-\pi\nu})e^{-\pi\nu/2}s^{-i\nu/2}+\\
  \sqrt{4\pi}(2s)^{i\nu/2-1}e^{is^2-\pi\nu/4-i\pi/4}(1-e^{-\pi\nu}).
\end{gather*}
Combining the asymptotics $\sim \exp(\pm is^2/2)$ at $s\rightarrow-\infty$ from subsections 3 and 4, we obtain the second LZSM matrix in~\eqref{lz-matrix} (lower signs).

\end{document}